\documentclass[twocolumn,showpacs,prl]{revtex4}
\usepackage{graphicx}
\usepackage{dcolumn}
\usepackage{amsmath}

\begin{document}
\title{ Noncommutative Quantum Cosmology}
\author{H. Garc\'{\i}a-Compe\'an$^{a}$, O. Obreg\'on$^{b}$ and C. Ram\'{\i}rez$^c$}
\affiliation{$^{a}$ {\it Departamento de F\'{\i}sica,\\
 Centro de Investigaci\'on y de Estudios Avanzados del IPN}\\
P.O. Box 14-740, 07000, M\'exico D.F., M\'exico\\
E-mail address: {\tt compean@fis.cinvestav.mx}\\
$^b$ {\it Instituto de F\'{\i}sica de la Universidad de Guanajuato}\\
P.O. Box E-143, 37150, Le\'on Gto., M\'exico\\
E-mail address: {\tt octavio@ifug3.ugto.mx}\\
$^c$ {\it Facultad de Ciencias F\'{\i}sico Matem\'aticas,\\
Universidad Aut\'onoma de Puebla}\\
P.O. Box 1364, 72000, Puebla, M\'exico\\
E-mail address: {\tt cramirez@fcfm.buap.mx}}
\date{\today}
\vskip -.5truecm
\begin{abstract}

We propose a model for noncommutative quantum cosmology by means of a deformation of minisuperspace.  For
the Kantowski-Sachs metric we are able to find the exact wave function. 
 We construct
wave packets and show that  new quantum states that ``compete'' to be the most probable state appear,
in clear contrast with the commutative case. A tunneling process could be possible among these states.

\end{abstract}
\maketitle
\vskip -.5truecm

The old proposal of noncommutativity in space-time \cite{snyder} has been recently subject of renewed
interest (see reviews \cite{connes,varilly,douglas}). This is a consequence of the developments in
M(atrix) theory \cite{banks} and string theory, from which noncommutativity arises in the low energy
effective field theory on a D-brane in a constant background B-field \cite{cds,ho,shomerus,sw}.

One of the most exciting recent applications of the idea of a minimal size to field theory, is that
concerning the description of Yang-Mills instantons in four dimensional noncommutative spacetimes. 
It has been shown that in these spaces instantons  acquire an effective size in terms of the
noncommutativity parameter $\theta$. As a consequence, the moduli space of noncommutative 
instantons no longer has the singularities corresponding to small instantons \cite{ns}. This
effect has a nice stringy interpretation \cite{sw}.

Noncommutative gravity has been considered in \cite{moffatone,moffattwo,cham}. In particular in reference
\cite{cham} a deformed Einstein gravity is constructed by using the Seiberg-Witten map \cite{sw}, gauging
the noncommutative ISO(3,1) group.  Some aspects of noncommutative $2+1$-dimensional Chern-Simons gravity
have been also studied \cite{chandia,gradi}.

One of the puzzles in quantum gravity is the measurement of length, which seems to be limited to
distances greater than the Planck length $L_P$, because to locate a particle we would need an
energy greater than the Planck mass $M_P$.  The corresponding gravitational field will have an horizon
given by the Schwarzschild radius $R= {2 G M_{P} \over c^2} = 2 L_{P}$ and, whatever happens inside,
this radius is shielded and therefore a minimal size should exist for quanta of space and time
configuration.

Consequently, at very early times of the universe, before the Planck time, nontrivial effects of
noncommutativity can be expected.

In the study of homogeneous universes, the metric depends only on the time parameter. Thus, the space
dependence can be integrated out in the action and a model with a finite dimensional configuration space
arises, called also minisuperspace, whose variables are the three-metric components. These theories have
been considered by themselves, and their quantization is performed following the rules of quantum
mechanics.

The minisuperspace construction is a procedure to define quantum cosmology models in the search to describe
the quantum behavior of the very early stages of the universe \cite{ryan,hh}. By defining these models one
necessarily
freezes out degrees of freedom, so that these are only simple and probably approximate models of 
full quantum gravity at Planckian times. Formally, a kind of Klein Gordon equation 
arises in these models, which describes the quantum behavior of the universe. 
Actually, the validity of this approach remains as an open
question to date. For example, within the context of Bianchi IX cosmology, it has been shown that imposing
additional symmetry on the model alters its physical predictions \cite{kr}.  
On the other hand, it has been
argued that one can find conditions that must be satisfied to justify the minisuperspace approximation
\cite{sinha}. In string theory formalism, general relativity, and consequently the Wheeler-DeWitt equation,
corresponds to the s-wave approximation \cite{susskind}.  Nevertheless , by considering a more general
analysis, \cite{halliwell}, it seems that we can expect
that the fundamental behavior of the wave function will be preserved.  

Recently, attempts to connect M(atrix)-string theory to cosmology on the brane \cite{ivonne} have been
done. The fact that in the former theory noncommutativity has been shown to be present, motivated us to
consider it also in models of the universe. In this paper we make a proposal in order to explore the
influence of noncommutativity at early times, by the introduction of an effective noncommutativity in quantum
cosmology.  Instead of a deformation of space-time, we consider a {\it ``deformation of
minisuperspace"}.

We will assume that the minisuperspace variables do not commute, as it has been proposed for the
spacetime coordinates. Our proposal is inspired in various related results in the literature, some of
them already referred here, as is the case of the Seiberg-Witten map \cite{sw}, where by demanding gauge
invariance, a redefinition of the gauge fields as an expansion on the noncommutativity parameters is
obtained. As a consequence of space-time noncommutativity, the fields do not commute among them in a
specific manner dictated by the Moyal product. In our ansatz, we propose a simple and direct
noncommutativity among certain components of the gravitational field. 

It is well known that already in
the usual spacetime, noncommutativity is usually defined in the ``preferred'' frame of cartesian
coordinates, where the noncommutative parameters are taken to be constant. For any other coordinate
systems, the corresponding noncommutativity will be, in general, in terms of parameters related in a
complicated manner to those of cartesian coordinates. For the even-dimensional case, the spacetimes can
be interpreted as symplectic manifolds with the noncommutativity parameters playing the role of a
symplectic form. In this case, Darboux's theorem ensures that always, locally, there exists a coordinate
system in which the components of the symplectic form are constant. In string theory this is assumed
when a `constant' $B$-field is considered.

On the other hand, the Seiberg-Witten map for gravitation has been proposed in \cite{cham}, where 
noncommutative tetrads and connections are computed. In the case of quantum cosmology, 
the minisuperspace variables play the role of the ``coordinates'' of the configuration space. Thus,
it seems reasonable to propose a kind of noncommutativity among these specific gravitational variables, 
as it is the case in standard spacetime, when cartesian coordinates are selected. 
From the canonical quantization of a cosmological model a quantum mechanical version, with a finite 
number of degrees of freedom, of the general Wheeler-de Witt equation for general relativity arises. 
In this way, quantum cosmology is usually understood as a quantum mechanical
model of the universe. We will then further assume, that it can be proceeded as in standard 
noncommutative quantum mechanics.

The noncommutativity we propose, can be reformulated in terms of a Moyal deformation of the Wheeler-DeWitt equation,
similar to the
case of the noncommutative Schr\"odinger equation \cite{ncqm}. Actually, cosmology depends only on time and
a Moyal product of functions of only one variable is trivially realized.  Other authors
have
worked out noncommutativity in the early universe, however without considering the gravitational field. In Ref.  
\cite{mazumdar} the noncommutativity of space-time is interpreted as a magnetic field on a horizon scale.
In \cite{cgs,lizzi}, it is argued that noncommutativity affecting matter or gauge fields could have played
an important role to produce inflation.

As an example of our proposal, we will consider the cosmological model of the Kantowski-Sachs metric. In
the parametrization due to Misner, this metric looks like \cite{misner}:

\begin{equation} ds^2 = - N^2 dt^2 + e^{2 \sqrt{3} \beta} dr^2 + e^{-2 \sqrt{3} \beta} e^{-2 \sqrt{3}
\Omega} (d \theta^2 + {\rm sin}^2 \theta d \varphi^2). 
\label{1} 
\end{equation} 
The corresponding Wheeler-DeWitt equation, in a particular factor ordering, can be written as

\begin{equation}
{\rm exp} (\sqrt{3} \beta +2 \sqrt{3} \Omega) [-P^2_\Omega +P^2_\beta
 - 48 {\rm exp} (-2 \sqrt{3} \Omega)] \psi (\Omega,\beta)=0, \label{2}
\end{equation}
where $P_\Omega = -i \frac{\partial}{\partial\Omega}$ and $P_\beta=-i 
\frac{\partial}{\partial\beta}$. Thus, in this parametrization the Wheeler-DeWitt equation 
has a simple form, which can be formally identified with usual quantum mechanics in cartesian coordinates.

The solutions to this Wheeler-DeWitt equation are given by  \cite{misner}

\begin{equation}
\psi_\nu^\pm(\beta,\Omega)= e^{\pm i\nu \sqrt{3} \beta} K_{i\nu}(4 e^{-\sqrt{3} \Omega}),
\label{3}
\end{equation}
where $K_{i\nu}$ is the modified Bessel function.
Packet waves of these solutions have been constructed as superpositions 
of these solutions. Summing over $e^{i \nu \sqrt{3} \beta}$ and 
$e^{-i \nu \sqrt{3} \beta}$ to make real trigonometric
functions, the ``Gaussian" state
\begin{eqnarray}
\Psi(\beta,\Omega)= 2i{\cal N}
\int_{0}^{\infty}\nu\left[\psi_\nu^+(\beta,\Omega)-\psi_\nu^-(\beta,\Omega)\right] d\nu&&\nonumber\\ = {\cal
N} e^{-\sqrt{3} \Omega} {\rm sin h} (\sqrt{3} \beta) {\rm exp} [-2\sqrt{3} e^{-\sqrt{3} \Omega} {\rm cosh}
(\sqrt{3} \beta)],&&
\label{4}
\end{eqnarray}
has been obtained \cite{misner,or}. A possible connection with quantum black
holes \cite{or,cavaone} and quantum wormholes \cite{cavatwo,garay} has been 
suggested.

For our noncommutative proposal of quantum cosmology, we will follow the procedure outlined above. We will
assume that the ``cartesian coordinates" $\Omega$ and $\beta$ of the Kantowski-Sachs minisuperspace obey a kind of
commutation relation, like the ones in noncommutative quantum mechanics \cite{ncqm}

\begin{equation}
[\Omega,\beta]=i\theta.
\end{equation}
As stated above, this is a particular ansatz in these configuration coordinates. 
The relation  with other minisuperspace coordinates would follow in a similar way  as in standard spacetime. 

As usual, this deformation can be reformulated in terms of a noncommutativity of 
minisuperspace functions, with the Moyal product,
\begin{equation}
f(\Omega,\beta)*g(\Omega,\beta)=f(\Omega,\beta)e^{i{\theta \over 2}( \overleftarrow{\partial}_\Omega
\overrightarrow{\partial}_\beta- \overleftarrow{\partial}_\beta \overrightarrow{\partial}_\Omega)}
g(\Omega,\beta). 
\end{equation}

Now, our noncommutative Wheeler-DeWitt equation will be

\begin{equation}
{\rm exp} (\sqrt{3} \beta + 2 \sqrt{3} \Omega) * [-P^2_\Omega 
+ P^2_\beta - 48 {\rm exp} (-2 \sqrt{3} \Omega)] * \psi (\Omega.
\beta)=0.
\label{7}
\end{equation}

Then, as is well known in noncommutative quantum mechanics \cite{ncqm}, the original phase-space, as well as its
symplectic structure, is modified. It is possible to reformulate it in terms of commutative variables
and the ordinary product of functions, if new variables are introduced,
$\Omega  \to \Omega - {1 \over 2 } \theta P_{\beta}$ and  $\beta \to \beta  - {1 \over
2 } \theta P_{\Omega}$, the momenta remain the same. As a consequence, the original equation changes, 
with a potential modified due to these new coordinates. 

\begin{equation} 
V(\Omega, \beta)*\psi(\Omega,\beta)=V(\Omega-\frac{1}{2}\theta P_{\beta}, \beta - {1 \over 2}
\theta P_{\Omega})\psi(\Omega,\beta).
\end{equation}
Thus, we get

\begin{equation}
[-\frac{\partial^2}{\partial\Omega^2} + \frac{\partial^2}{\partial\beta^2} + 
48 {\rm exp} (-2 \sqrt{3} 
\Omega + \sqrt{3} \theta P_\beta)] \psi (\Omega, \beta) = 0. 
\label{9}
\end{equation}
Assuming a separation of variables with the ansatz 

\begin{equation}
\psi (\Omega, \beta) = {\rm exp} (\sqrt{3}\nu\beta) \chi (\Omega), 
\label{10}
\end{equation}
we observe that the operator $P_\beta$ in the exponential in (9) 
will shift the wave function by a factor 

\begin{equation}
\psi (\Omega, \beta - i \sqrt{3} \theta) = {\rm exp} (-3i\nu \theta)
\psi (\Omega, \beta), 
\label{11}
\end{equation}
thus $\chi(\Omega)$ must satisfy the equation

\begin{equation}
\left[- \frac{d^2}{d\Omega^2} + 48 {\rm exp} (-3i\nu \theta) {\rm exp} 
(- 2 \sqrt{3} \Omega) + 3\nu^2\right] \chi (\Omega) = 0. 
\label{12}
\end{equation}
This equation can be solved, in the same manner as in the commutative 
case, by a modified Bessel function $K_{i\nu}$. Therefore the wave function will be given by 

\begin{equation}
\psi_\nu^\pm \left(\Omega, \beta \right) = e^{\pm i \sqrt{3} \nu \beta} 
K_{i\nu} \left\{ 4 {\rm exp} 
\left[ - \sqrt{3} \left ( \Omega \mp \frac{\sqrt{3}}{2} \nu \theta 
\right) \right] \right\}. 
\label{13}
\end{equation}

These are the solutions to the Wheeler-DeWitt equation (9). Note that noncommutativity induces a difference
of the arguments of the Bessel functions in these solutions. Moreover, from its form, 
we can expect that the noncommutativity
effects are enhanced for $\psi^+$.  Thus, for the particular model we have chosen, 
the solution (13) allows an exact analysis, without the need of a $\theta$ expansion.

In order to see the consequences of noncommutativity, let us consider a wave packet weighted by a
Gaussian,

\begin{equation}
\Psi(\Omega,\beta)= {\cal N}\int_{-\infty}^{\infty}e^{-a(\nu-b)^2} \psi_\nu^{+} \left(\Omega, \beta \right)
d\nu.
\end{equation}
This integral is performed numerically. We are interested to see which is the influence of the
$\theta$ parameter, but we are as well interested on consequences for the values of 
the $\Omega$ and $\beta$ variables, which in particular can provide information about the anisotropy. 

As mentioned, a minimal size should exist for quanta of space and 
time configuration, and we are considering the very early time of the 
universe, where the influence  of noncommutativity could have played a
role in its quantum behavior.  The figures 1, 2 and 3, computed for the values 
$a=1.5$ and $b=1.3$, show the dramatic changes that the universe 
could have had if noncommutativity was present.

Figure 1 corresponds to $\theta=0$, the standard commutative  case, and is presented for $\Omega$ 
in the range  [0,10] and $\beta$ in the range [-10,10], it shows {\it only one} preferred state of the 
universe around $\beta=0$. For $\theta=4$, things have drastically changed, more peaks appear 
that seem {\it to compete to be the most probable state} of the universe. 
The peaks are no more around only of $\beta=0$, but they appear, symmetrically, 
around other values of $\beta$. Moreover, the original peak has changed its $\Omega$ value. 
Furthermore, figure 3 shows more clearly the existence of the possibility of tunneling among these states.

Noncommutativity in minisuperpace creates then, new possible states 
of the universe.  So the universe we live today could have evolved  not only 
from  the state for the commutative case (figure 1), but from any of the other 
states of the noncommutative model, like those of figure 2. It could have jumped, 
by means of a tunneling process, as figure 3 
shows more clearly, from one universe (one state) to other universe 
(other state). 

\begin{figure}
\begin{center}
\includegraphics[width=8 cm]{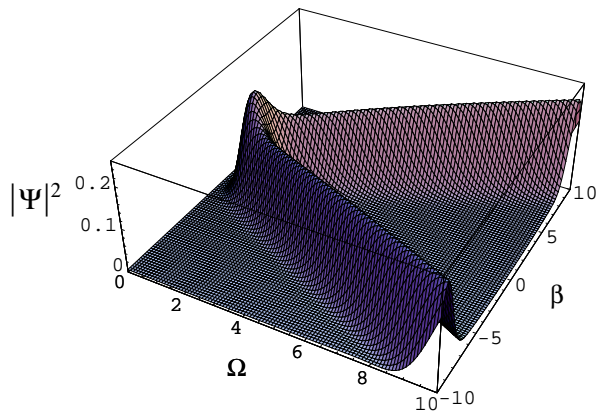}\\
\end{center}
\caption{Variation of $|\Psi|^2$ with respect to $\Omega$ and $\beta$, at the value 
$\theta=0$. It shows only one possible universe around $\Omega=4.812$ and $\beta=0$.}
\label{wavef1}
\end{figure}

\begin{figure}
\begin{center}
\includegraphics[width=8 cm]{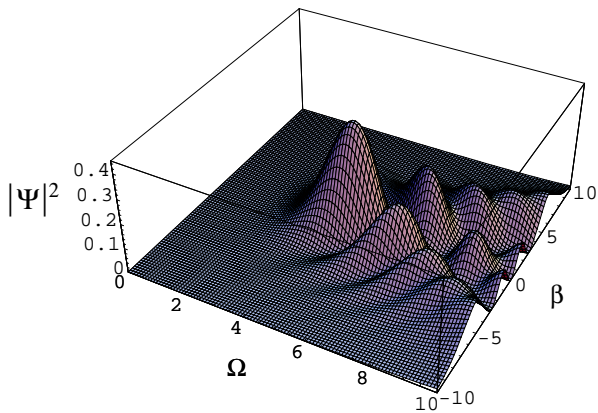}\\
\end{center}
\caption{Variation of $|\Psi|^2$ with respect to $\Omega$ and $\beta$, at the value 
$\theta=4$. It shows many peaks corresponding to different possible universes.}
\label{wavef2}
\end{figure}

\begin{figure}
\begin{center}
\includegraphics[width=7 cm]{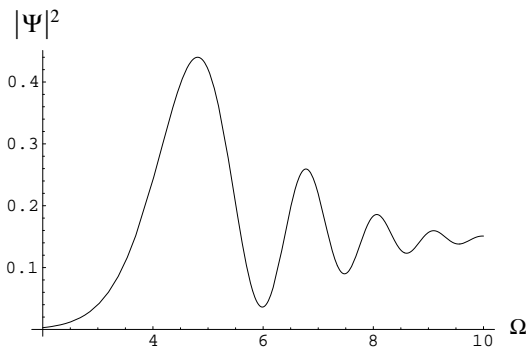}\\
\end{center}
\caption{
Variation of $|\Psi|^2$ for $\theta=4$, along the line of the main peaks in figure 2. 
It shows the tunneling possibility among these states.}
\label{wavef3}
\end{figure}

Although quantum cosmology, as discussed above, is only a limited model in an attempt to describe
some of the features of the quantum theory of the universe, the consideration of noncommutativity seems to
be one way to take into account the presence of constant Neveu-Schwarz background $B$-fields in M(atrix)
\cite{banks} and string theory, at early times in the universe.

As already mentioned, there are in the literature proposals for a noncommutative theory of gravity
\cite{moffatone,moffattwo,cham}.  In particular, in reference \cite{cham} the Seiberg-Witten map was
used to construct a deformed Einstein gravity. By means of this result, one could try to find the
corresponding noncommutative Wheeler-DeWitt equation for specific cosmological models. In that case the
$\theta$ terms corresponding to spacetime noncommutativity could be considered in order to search for
another way to define a noncommutative cosmological model, even though cosmology depends only on time.
The computation of this Wheeler-DeWitt equation is a very complicated task, because at each $\theta$
order higher derivative terms will appear. If such a noncommutative Hamiltonian model could be defined,
its quantum cosmological solutions and the corresponding states could be compared with those obtained by
means of our noncommutative minisuperspace proposal. In this context the $\theta$ parameter we have proposed
(5), could be a kind of effective noncommutative parameter.  Such an approach is currently under
exploration.

Further work in more realistic cosmological models, including matter, will be needed to search for 
constraints on the range of values of the $\theta$ parameter in the early stages of evolution of 
the universe. In our proposal this is in principle possible, because it means to enlarge the
minisuperspace configuration space and then consider an appropriate commutativity among its coordinates. Following
these lines, we will consider also the supersymmetric extension \cite{susyqc} to these models.

In previous works, the Wheeler-DeWitt equation of the Kantowski-Sachs model has been related to quantum black
holes \cite{or,cavaone} and wormholes \cite{cavatwo,garay}. It could be interesting to search for an
extension of our results to possible noncommutative versions of these quantum gravitational systems.

Our simple proposal provides a picture of the dramatic influence that noncommutativity could have played
at early stages of the universe.  We have been able to obtain and work with an exact quantum solution for
the Kantowski-Sachs metric, without the need to expand on the $\theta$ 
parameter. The corresponding  wave packet exhibits new stable states of the
universe with similar probabilities, showing peaks  around 
$\beta$ values different from zero.  A tunneling process could happen 
between these states. By these means there are different possible universes 
(states) from which our present universe could have evolved and also could have tunneled in the past, 
from one universe (state) to other one. Further
work is needed to analyze other physical consequences of this and other 
 more realistic quantum cosmological models,
taking into account the influence of matter. It will be also necessary to 
extend these
ideas to other, more general gravitational models. In particular, it would be very interesting to reinterpret
the results of reference \cite{tsagas}, concerning the influence of a primordial magnetic field in classical
cosmology. The corresponding quantum model should be constructed and compared with our proposal 
in terms of the noncommutativity of the minisuperspace.
Results in these directions and those mentioned above will be presented elsewhere.

\vskip 1truecm

\centerline{\bf Acknowledgments}
This work was supported in part by CONACyT Mexico Grant Nos. 28454E and 33951E.

\end{document}